\documentclass[12pt]{article}
\usepackage{epsfig}

\def\beq{\begin{equation}}
\def\eeq#1{\label{#1}\end{equation}}
\def\eeqn{\end{equation}}

\def\beqa{\begin{eqnarray}}
\def\eeqa#1{\label{#1}\end{eqnarray}}
\def\eeqan{\end{eqnarray}}

\let\bar=\overbar

\def\Dslash{\not{\hbox{\kern-4pt $D$}}}
\def\dslash{\not{\hbox{\kern-2pt $\del$}}}

\def\msb{{\bar{\ssstyle M \kern -1pt S}}}

\def\Title#1{\begin{center} {\Large {\bf #1} } \end{center}}

\begin{document}

\noindent
PITHA 02/10\\
hep-ph/0207228\\
July 18, 2002\\

\Title{Recent results from the QCD factorization approach to 
non-leptonic $B$ decays\footnote{
Talk presented at Flavor Physics and CP Violation (FPCP), Philadelphia, 
U.S.A., 16-18 May 2002.}}

\bigskip\bigskip

\begin{raggedright}  

{\it M. Beneke\\
Institut f\"ur Theoretische Physik E\\
RWTH Aachen\\
Sommerfeldstr. 28\\
D-52074 Aachen, GERMANY}
\bigskip\bigskip
\end{raggedright}

\section{Introduction}

In this talk I report recent results on hadronic $B$ decays obtained 
with the QCD factorization approach. In the first part I update the 
fit of the Wolfenstein parameters $(\bar\rho,\bar\eta)$ to  
CP-averaged $B\to\pi\pi, \pi K$ branching fractions performed in 
\cite{BBNS3} to account for the new experimental data, and give the 
correlation between the time-dependent and direct CP asymmetry in 
$B_d\to \pi^+\pi^-$ decay, also based on the calculation of 
\cite{BBNS3}. In the second part I present an investigation of 
$B$ decays into final states containing an $\eta$ or $\eta^\prime$ 
meson \cite{MBMN}. I discuss the new theoretical issues that arise 
for flavour singlet mesons and present preliminary numerical results 
that appear to reproduce the pattern of experimental data reasonably 
well. 

The QCD factorization approach \cite{BBNS1} uses heavy 
quark expansion methods ($m_b\gg\Lambda_{\rm QCD}$) and 
soft-collinear factorization (particle energies $\gg\Lambda_{\rm QCD}$) 
to compute the matrix elements 
$\langle f|O_i|\bar B\rangle$ relevant to hadronic $B$ decays 
in an expansion in $1/m_b$ and 
$\alpha_s$. Only the leading term in $1/m_b$ assumes a simple 
form. The basic formula is 
\begin{eqnarray}
\label{fact}
\langle M_1 M_2|O_i|\bar B\rangle 
&\hspace*{-0.2cm}=&\hspace*{-0.2cm} 
F^{B\to M_1}(0)\int_0^1 \!\!du\,T^I(u)
\Phi_{M_2}(u) \nonumber\\[0.0cm]
&&\hspace*{-2.8cm}
+\!\int \!d\xi du dv \,T^{II}(\xi,u,v)\,\Phi_B(\xi)\Phi_{M_1}(v) 
\Phi_{M_2}(u),
\end{eqnarray}
where $F^{B\to M_1}$ is a (non-perturbative) form factor, 
$\Phi_{M_i}$ and $\Phi_B$ are light-cone distribution 
amplitudes and $T^{I,II}$ are perturbatively calculable 
hard scattering kernels. Although not strictly proven to all orders 
in perturbation theory, the formula is presumed to be 
valid when both final state mesons are light. ($M_1$ is 
the meson that picks up the spectator quark from the $B$ meson.) The formula 
shows that there is no long-distance interaction between the
constituents of the meson $M_2$ and the $(B M_1)$ system at leading 
order in $1/m_b$. This is the precise meaning of factorization. 
A summary of results that have been obtained in the QCD factorization 
approach is given in \cite{KEK}. 

Factorization is not 
expected to hold at subleading order in $1/m_b$. 
Attempts to compute subleading 
power corrections to hard spectator-scattering in perturbation theory 
usually result in infrared divergences, which signal the breakdown 
of factorization. Some power corrections related to scalar currents 
are enhanced by factors such as 
$m_\pi^2/((m_u+m_d)\Lambda_{\rm QCD})$ \cite{BBNS1}. 
At least these effects should be estimated and included into the 
error budget. All weak 
annihilation contributions belong to this class of effects.

\section{CP-averaged $B\to\pi\pi, \pi K$ branching fractions}

The possibility to determine the CP-violating angle $\gamma$ 
by comparing the calculation of branching fractions 
into $\pi\pi$ and $\pi K$ final states with the corresponding 
data has been investigated in detail \cite{BBNS3} 
(see also \cite{Du:2001hr}). The branching fractions for the modes 
$B^+\to\pi^+\pi^0$
  and $B^+\to\pi^+ K^0$, which depend only on a single weak
  phase to very good approximation, are well described by the 
theory. This demonstrates that the
  magnitude of the tree and penguin amplitude is obtained
  correctly, where for the penguin amplitude the 1-loop radiative 
correction is important to reach this conclusion. There is, however, a 
relatively large normalization 
  uncertainty for the $\pi K$ final states, which are sensitive to 
  weak annihilation and the strange quark mass through the scalar
  penguin amplitude. This uncertainty can be partially eliminated by
  taking ratios of branching fractions. The agreement is less 
  good for branching fractions with significant interference of tree
  and penguin amplitudes, if $\gamma$ is assumed to take values 
  around $55^\circ$ as favoured by indirect constraints. 

\begin{figure}[p]
\begin{center}
\hspace*{-7.5cm}
\epsfig{file=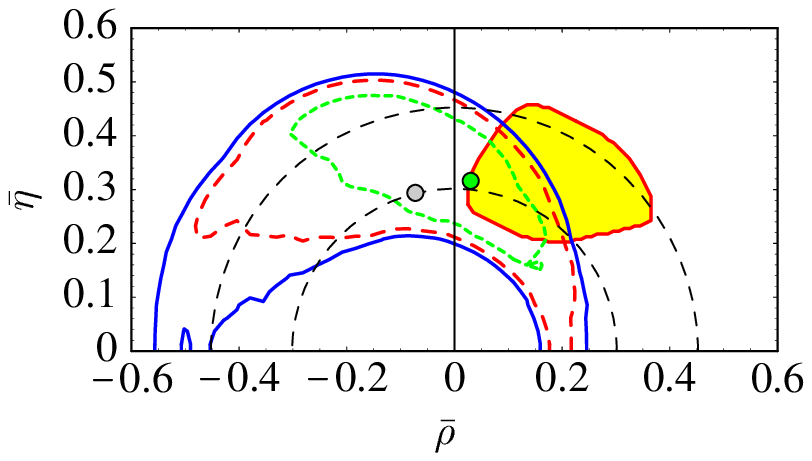,height=1.7in}
\vskip-1.74in
\hspace*{4.8cm}
\epsfig{file=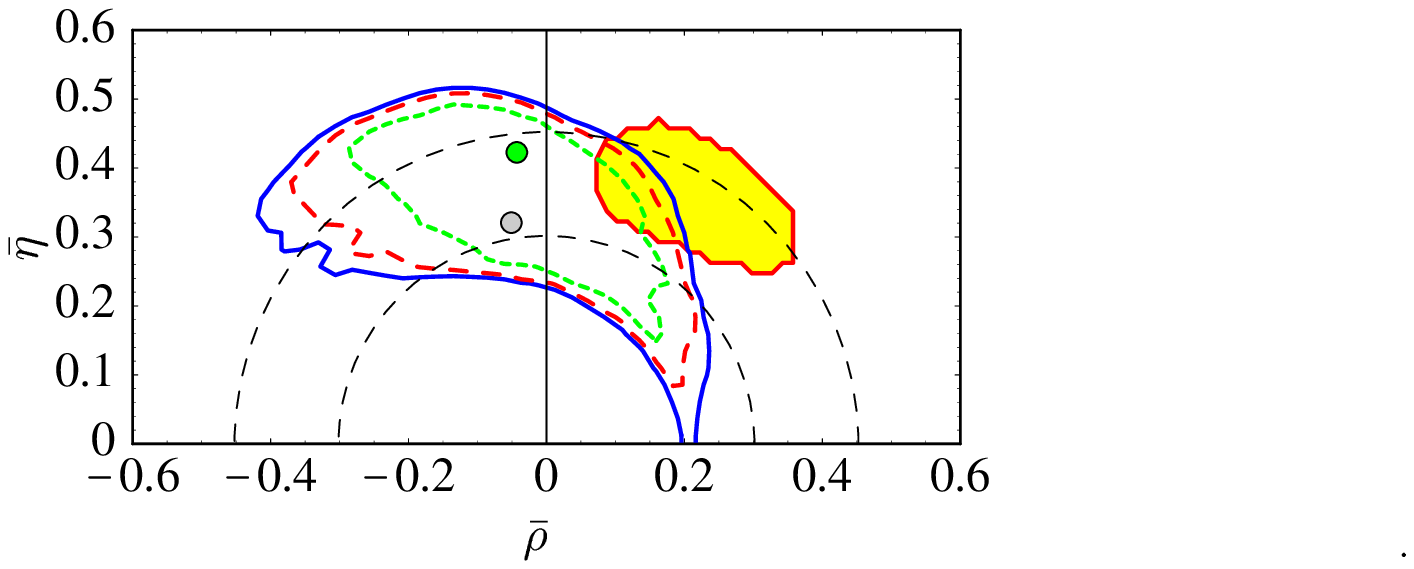,height=1.7in}
\caption{95\% (solid), 90\% (dashed) and 68\% (short-dashed) confidence level 
contours in the $(\bar\rho,\bar\eta)$ plane obtained from a global 
fit to the CP averaged $B\to\pi K,\pi\pi$ branching fractions, using 
the scanning method as described in 
\cite{Hocker:2001xe}. The darker dot shows the
overall best fit, whereas the lighter dot indicates the best fit for the
default hadronic parameter set. The left panel reproduces the fit 
of \cite{BBNS3} reflecting the status of spring 2001; the right panel 
summarizes the current results. 
The light-shaded region indicates the region 
preferred by the standard global fit \cite{Hocker:2001xe}, 
excluding (including)  the direct measurement of $\sin(2\beta)$ 
in the left (right) panel.}
\label{fig:pik}
\end{center}
\end{figure}
\begin{table}[p]
\begin{center}  
\vskip0cm
\begin{tabular}{l|c|c|c}
 Decay Mode & Exp. Average\hspace{0.15cm}  & 
\hspace*{0.1cm}  Default fit \hspace*{0.1cm} & \hspace*{0.4cm}  Fit2 
\hspace*{0.4cm}\\[0.1cm]
\hline
&&& \\[-0.3cm]
$B^0\to\pi^+\pi^-$& $5.15\pm 0.61$  & 5.12 & 5.24\\
$B^\pm\to\pi^\pm\pi^0$ &$4.88\pm 1.06$  & 5.00 & 4.57 \\
$B^0\to\pi^0\pi^0$ &$-$  & 0.78 & 0.94 \\[0.1cm]
\hline
&&& \\[-0.3cm]
$B^0\to\pi^\mp K^\pm$ & $18.56\pm 1.08$  & 17.99 & 18.47 \\
$B^\pm\to\pi^0 K^\pm$ & $11.49\pm 1.26$  & 12.07 & 11.83 \\
$B^\pm\to\pi^\pm K^0$ & $17.93\pm 1.70$  & 15.65 & 17.88 \\
$B^0\to\pi^0 K^0$     & $8.82\pm 2.20$   & 5.55  & 6.87 \\[0.1cm]
\hline
\end{tabular}
\caption{\label{tab1} CP-averaged $B\to\pi\pi, \pi K$ branching
  fractions (in units of $10^{-6}$): 
data vs. results from the fit. The default fit to
$(\bar\rho,\bar\eta)$ (returning $|V_{ub}/V_{cb}|=0.085$, 
$\gamma=116^\circ$ with $\chi^2 =4.5$) refers to the 
default theory parameter set as explained in the text. 
``Fit2'' (returning $|V_{ub}/V_{cb}|=0.079$, 
$\gamma=97^\circ$, $\chi^2 =1.0$) refers to a fit without
annihilation contributions and chirally enhanced spectator corrections
but with $m_s=80\,$MeV [$100\,$MeV], $\lambda_B=200\,$MeV [$350\,$MeV]
and $R_{\pi K}=0.8$ [0.9] (default values in square brackets). The
experimental average is based on $9\,\mbox{fb}^{-1}$ from CLEO, 
$29.1\,\mbox{fb}^{-1}$ from Belle and $55.6\,\mbox{fb}^{-1}$ from
Babar~\cite{bartoldus}.
}
\end{center}
\end{table}

In \cite{BBNS3} a fit of the Wolfenstein parameters 
$(\bar\rho,\bar\eta)$ to the six measured CP-averaged 
$B\to\pi\pi, \pi K$ branching fractions has been performed. 
The result of this fit is 
shown in the left panel of Figure~\ref{fig:pik}. (The details of the
fit procedure can be found in \cite{BBNS3}). We now repeat this fit 
with the new world averages as presented at this conference 
\cite{bartoldus}, see Table~\ref{tab1}. There are no dramatic changes 
in the data since spring 2001, but the small shifts of the various
branching fractions (for instance, in the 
final state $\pi^0 K^0$) all work towards better agreement with the 
theoretical calculation, resulting in an improved fit. (The best fits 
with theory parameters in the allowed ranges have $\chi^2\approx 0.5$.) On the
theoretical side we changed the allowed values of the strange quark
mass and  the $B$ meson decay constant 
to $[75,125]\,\mbox{MeV}$ (from $[85,135]\,\mbox{MeV}$) and 
$[170,230]\,\mbox{MeV}$ (from $[140,220]\,\mbox{MeV}$), respectively, 
to account for a change in the theoretically favoured ranges. The
result of the current fit is shown in the right panel of 
Figure~\ref{fig:pik}. The last two columns of Table~\ref{tab1} give 
the fitted branching fractions for the default theory parameter set 
(corresponding to the parameters used in \cite{BBNS3} and the central
values of the new ranges for $m_s$ and $f_B$) and a second set, where 
all annihilation effects and chirally enhanced spectator interactions 
are switched off. The second set therefore shows that very good fits 
can also be obtained without these theoretically uncertain
power-suppressed effects. While a large range of values of $\gamma$ 
remains compatible with data, and the result of the fit is 
consistent with the standard fit 
based on meson mixing and $|V_{ub}|$, it shows a preference for 
$\gamma$ near $90^\circ$, or, for smaller $\gamma$, smaller $|V_{ub}|$.

\section{CP asymmetries in $B_d\to\pi^+\pi^-$ decay}

\begin{table}[b]
\begin{center}
\begin{tabular}{l|cc}  
Experiment &  $S_{\pi\pi}$ &  $C_{\pi\pi}$ \\ \hline
Babar  &   $-0.01\pm 0.37 \pm 0.17$   &  $-0.02\pm 0.29 \pm 0.09$    \\
Belle  &   $-1.21^{+0.38+0.16}_{-0.27-0.13}$    &   
$-0.94^{+0.31}_{-0.25}\pm 0.09$       \\ \hline
\end{tabular}
\caption{CP asymmetries in $B_d\to\pi^+\pi^-$ from Babar and Belle 
\cite{alpha1,alpha2}.}
\label{tab:pipi}
\end{center}
\end{table}

The QCD factorization approach allows us to interpret directly the 
mixing-induced and direct CP asymmetry in $B_d\to\pi^+\pi^-$ decay without 
resort to other decay modes, 
since the tree and penguin amplitudes are both computed. 
The time-dependent asymmetry is defined by
\begin{eqnarray}
   A_{\rm CP}^{\pi\pi}(t)
   &=& \frac{\mbox{Br}(B^0(t)\to\pi^+\pi^-)
             - \mbox{Br}(\bar B^0(t)\to\pi^+\pi^-)}
            {\mbox{Br}(B^0(t)\to\pi^+\pi^-)
             + \mbox{Br}(\bar B^0(t)\to\pi^+\pi^-)} \nonumber\\[0.2cm]
   &=& - S_{\pi\pi} \sin(\Delta m_B\,t)
    + C_{\pi\pi} \cos(\Delta m_B\,t) ,
\end{eqnarray}
where $S_{\pi\pi} = \sin(2\alpha)$, if the penguin amplitude were zero,  
and $C_{\pi\pi}$ is the direct CP asymmetry. (This 
convention is related to those used by Babar and Belle by 
$S_{\pi\pi}=S_{\pi\pi}^{\rm Babar}=S_{\pi\pi}^{\rm Belle}$ and 
$C_{\pi\pi}=C_{\pi\pi}^{\rm Babar}=-A_{\pi\pi}^{\rm Belle}$.)

\begin{figure}[t]
\begin{center}
\hspace*{-7.6cm}
\epsfig{file=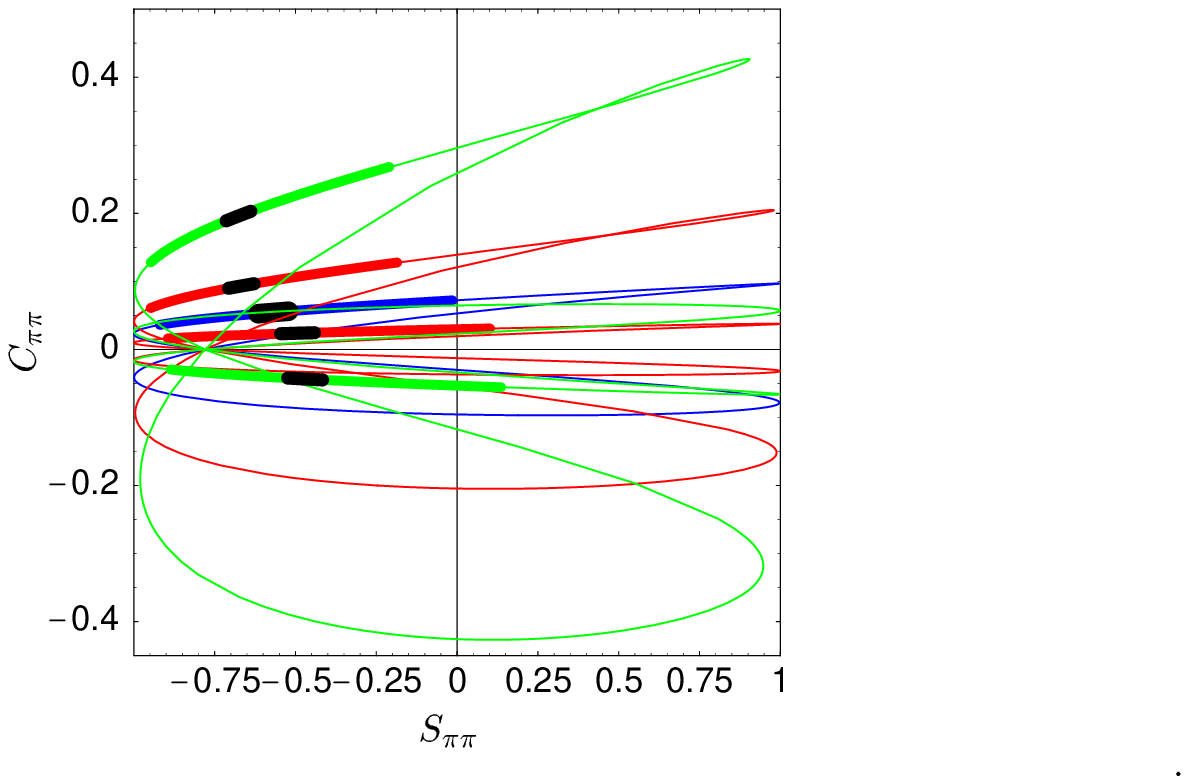,height=2.8in}
\vskip-7.15cm
\hspace*{4.3cm}
\epsfig{file=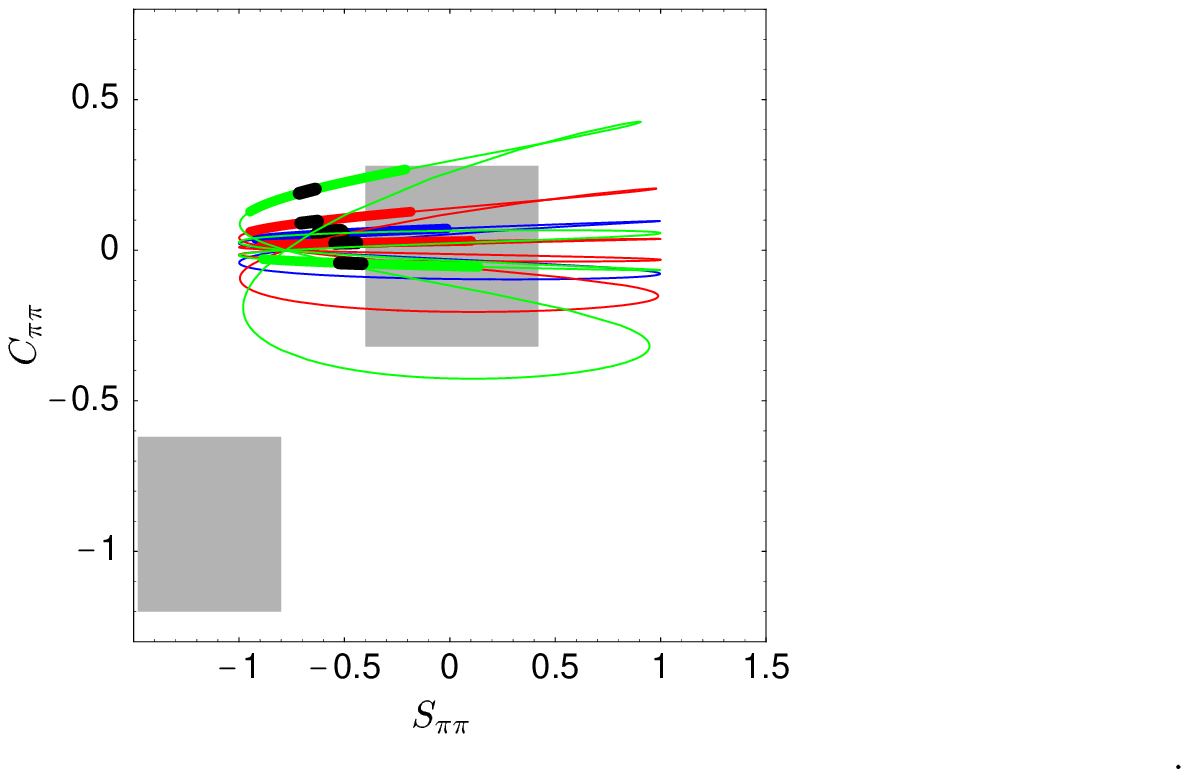,height=2.89in}
\caption{Predicted correlation between mixing-induced and direct CP 
asymmetry in $B_d\to\pi^+\pi^-$ decay. See text for explanation of the 
different curves. The right plot is a scaled version of the left plot, 
and includes the Belle result (upper left) and Babar 
result (center) 
with their 1$\sigma$ errors (gray rectangles). (Note, however, that 
the physical range is $|S_{\pi\pi}|^2+|C_{\pi\pi}|^2\leq 1$.)}
\label{fig:pipi}
\end{center}
\end{figure}

In \cite{BBNS3} we have shown how a measurement of $S_{\pi\pi}$ translates 
into a determination of $\sin(2\alpha)$ and results in a stringent 
constraint in the $(\bar\rho,\bar\eta)$ plane. In view of the new 
measurements of $S_{\pi\pi}$ and $C_{\pi\pi}$ from Babar 
\cite{alpha1} and Belle \cite{alpha2} 
(summarized in Table~\ref{tab:pipi}), it is also interesting to exhibit 
the correlation between the two observables. This is shown in 
Figure~\ref{fig:pipi}, where the $B\bar B$ mixing phase has been fixed such 
that $\sin(2\beta)=0.78$. We assume this phase to be experimentally given 
and do not require that $B\bar B$ mixing is described by the Standard Model. 
Each closed curve is then generated by specifying the theory input and  
letting $\gamma$ vary from 0 to $360^\circ$. The central (dark) curve 
refers to the calculation of $P/T$, the penguin-to-tree ratio, with 
the default theory parameter set, the two neighboring (lighter) curves 
refer to $P/T$ plus-minus its theoretical error without the error from 
weak annihilation (but including the one from $|V_{ub}|$), and the final 
(lightest) curves also include the error from weak annihilation. The black 
part on each curve marks the point $\gamma=60^\circ$; the fat line 
segment marks the range $[40^\circ,80^\circ]$ favoured by the standard 
unitarity triangle fit with larger $\gamma$ to the right of the black 
part. We can see from this that within the $1\sigma$ errors 
only the Babar result is compatible with the theoretical calculation, 
and it favours $\gamma$ somewhat larger than the standard 
unitarity triangle fit, but consistent with the CP-averaged 
branching fraction data discussed above.

\section{Final states containing $\eta^{(\prime)}$}

These final states are interesting because the available data exhibit 
an interesting pattern in $\Delta S=1$ decays (all branching fractions 
in units of $10^{-6}$):
\begin{eqnarray}
&& \hspace*{-1cm} 
\mbox{Br}(K\eta^\prime) \sim 70 \gg \mbox{Br}(K\eta) \sim 5 \,(?)
\nonumber\\[0.1cm]
&& \hspace*{-1cm}
\mbox{Br}(K^*\eta^\prime) \mbox{ not observed} <(?) 
\mbox{ Br}(K^*\eta) \sim 20
\end{eqnarray}
These results are difficult to account for in the naive factorization 
approach \cite{Ali:1997ex} (see also Table~\ref{tab:etak}).

The calculation of the flavour-singlet decay amplitude in QCD 
factorization involves several aspects specific to singlet mesons \cite{MBMN}. 
One of them is related to $\eta$-$\eta^\prime$ mixing for which we 
use the scheme advocated in \cite{FKS}, which amounts to assuming 
that mixing is described by a single mixing angle common to all 
matrix elements in the so-called quark-flavour basis. The other
aspects  
are related to the pure gluon content of singlet mesons, which leads 
to the following new effects:
\begin{enumerate}
\item The CKM-enhanced $b\to c\bar c s$ transition can contribute 
to singlet mesons by closing the charm lines and attaching two gluons 
to them. This effect amounts to assigning a charm decay constant 
to $\eta^{(\prime)}$, which by explicit calculation of the diagrams 
is found to be
\begin{equation}
f_P^c = -\frac{m_P^2}{12 m_c^2}\,\frac{f_P^q}{\sqrt{2}} 
\approx -2.5\,\mbox{MeV}\,[\eta^\prime],\,-1\,\mbox{MeV}\,[\eta],
\end{equation}
in agreement with \cite{Franz:2000ee}. 
Here $f_P^q$ is the up-quark decay constant 
and the result is obtained for $m_c\gg \Lambda_{\rm QCD}$. Note the 
absence of any factors of $\alpha_s$.
\item The singlet meson can be produced in a two-gluon state, where  
one gluon originates from a penguin $b\to s$ transition and the other 
from the spectator quark. We find a leading-power contribution from 
the configuration where the second gluon is soft, which implies that 
factorization breaks down in the conventional sense. Despite this, 
this non-perturbative contribution can be parameterized by a non-local 
$B\to K^{(*)}$ form factor, which introduces one new non-perturbative 
parameter. This effect is proportional to $\alpha_s C_{8g}$, where 
$C_{8g}$ is the Wilson coefficient of the chromomagnetic dipole operator, 
and can amount to several percent of the amplitude for $\eta^{(\prime)}$ 
mesons. 
\item There exists a singlet annihilation amplitude which is 
not power-suppressed in $m_b$, where two gluons radiate from the 
spectator quark and form an $\eta^{(\prime)}$ meson. 
\end{enumerate}

\begin{table}[t]
\begin{center}
\hspace*{-0.0cm}
\begin{tabular}{l|c|c|c}
Mode & Naive Fact.
 & QCD Fact. 
 & Exp. average \\[0.1cm]
\hline
&&& \\[-0.3cm]
$B^-\to K^-\eta'$ & 
$13$ & $ 47^{\,+40}_{\,-19}$
 & $75.1 \pm 6.2$ \\
$\bar B^0\to\bar K^0\eta'$ & 
$14$ & $47_{\,-19}^{\,+38}$
 & $61.0 \pm 12.5$ \\
$B^-\to K^-\eta$ & 
$0.7$ &  $1.3_{\,-0.8}^{\,+1.4}$ & $5.3\pm 1.8$ \\
$\bar B^0\to\bar K^0\eta$ &
0.1 &  $0.5_{\,-0.5}^{\,+1.0}$ &  $<9.3$   \\
$B^-\to K^-\pi^0$ &
$4.4$ & $9.4_{\,-3.4}^{\,+7.3}$  & $11.5\pm 1.3$ \\
$\bar B^0\to\bar K^0\pi^0$ & 
$2.2$ &  $6.4_{\,-2.8}^{\,+6.1}$ & $8.8\pm 2.2$\\[0.1cm]
\hline
&&& \\[-0.3cm]
$B^-\to K^{*-}\eta'$ &
$2.9$ & $3.3_{\,-3.3}^{\,+8.7}$  &  $<35$\\
$\bar B^0\to\bar K^{*0}\eta'$ &
$1.6$ & $2.1_{\,-2.1}^{\,+7.4}$  & $<24$ \\
$B^-\to K^{*-}\eta$ & 
$3.8$ & $9.3_{\,-6.1}^{\,+16.6}$  & $25.4 \pm 5.3$ \\
$\bar B^0\to\bar K^{*0}\eta$ & 
$4.3$ & $10.4_{\,-6.5}^{\,+17.3}$ & $16.4 \pm 3.0$ \\
$B^-\to K^{*-}\pi^0$ & 
$1.7$ &  $3.0_{\,-1.4}^{\,+4.0}$ & --  \\
$\bar B^0\to\bar K^{*0}\pi^0$ & 
$0.2$ & $0.8_{\,-0.6}^{\,+2.5}$ &  --  \\[0.1cm]
\hline
\end{tabular}
\caption{CP-averaged $B\to K^{(*)} (\eta^{(')},\pi^0)$ branching fractions 
in units of $10^{-6}$ 
in naive factorization and QCD factorization compared to experimental 
averages. Theoretical results are preliminary as explained in the text. All 
theoretical errors are strongly correlated.}
\label{tab:etak}
\end{center}
\end{table}

In Table~\ref{tab:etak} we present our (still preliminary) results 
for the CP-averaged $B\to K^{(*)} (\eta^{(')},\pi^0)$ branching fractions, 
not including effects 2. and 3. discussed above. We also used 
$|V_{ub}/V_{cb}|=0.09$ and $\gamma=60^\circ$ for the numerical 
evaluation. The dominant theoretical error is from the strange quark 
mass and weak annihilation and is correlated among the decay modes 
displayed in the Table. 

Despite the shortcomings of the above analysis (some of which we hope 
to rectify in the final publication \cite{MBMN}), we see that the QCD 
factorization calculation appears to reproduce the pattern of the 
data reasonably well within the uncertainties of the calculation,
which are very large for some of the decay modes.  
The basic features of this pattern can be understood 
from the structure of the penguin contributions to the decay amplitudes:
\begin{eqnarray}
A(\bar K\pi^0) &\sim& F^{B\to\pi}\,\frac{f_K}{\sqrt{2}}\,
\left(a_4^c(\pi K)+r_\chi^K a_6^c(\pi K)\right),
\nonumber\\
A(\bar K P) &\sim& F^{B\to P}\,\frac{f_K}{\sqrt{2}}\,
\left(a_4^c(P K)+r_\chi^K a_6^c(P K)\right) \hspace*{3.15cm} \mbox{(I)}
\nonumber\\
&&+\, F^{B\to K}\,\bigg(\left(\sqrt{2} f_P^q+f_P^s\right)\,
\left(a_3^c(KP)-a_5^c(KP)\right)\qquad (\mbox{small})
\nonumber\\
&&\hspace*{1.5cm}
+\, f_P^s\,
\left(a_4^c(KP)+r_\chi^P a_6^c(KP)\right)\bigg), \hspace*{2.2cm} \mbox{(II)}
\end{eqnarray}
where $P=\eta,\eta^\prime$ and $K=K,K^*$. For the $K\eta^\prime$ decay 
the two penguin amplitudes I and II add constructively enhancing the 
branching fraction by a large factor compared to $K\pi^0$. For 
$K\eta$, on the other hand, the two amplitudes nearly cancel since 
the strange decay constant in the $\eta$ satisfies $f_\eta^s/f_K
\sim -2/3$. Replacing $K$ by the vector meson $K^*$ again changes 
the pattern, because the scalar penguin amplitude $r_\chi^{M_2} 
a_6^c(M_1 M_2)$ changes sign for $M_1 M_2=K^*P$ (changing the sign of 
the term II) compared to $KP$ 
and becomes small for $M_1 M_2=PK^*$. As a consequence 
the terms I and II interfere destructively for $K^*\eta^\prime$ but 
constructively for $K^*\eta$, opposite to the case of a pseudoscalar 
kaon. These features are not different from the expectation in naive 
factorization. However, radiative corrections in QCD factorization 
enhance the penguin amplitudes significantly and improve the comparison 
with data. While it appears unlikely that one can obtain an accurate 
theoretical description of final states with singlet mesons 
from first principles, the present results clearly demonstrate the 
relevance of factorization to this class of charmless $B$ decays.

\bigskip\noindent 
I am grateful to M. Neubert, G. Buchalla and C.T. Sachrajda for 
collaboration on the topics discussed in this talk. I thank U. Nierste
for reading the manuscript.

\end{document}